# MoonGen: A Scriptable High-Speed Packet Generator

Paul Emmerich, Sebastian Gallenmüller, Daniel Raumer, Florian Wohlfart, and Georg Carle
Technische Universität München
Department of Computer Science
Chair for Network Architectures and Services
{emmericp|gallenmu|raumer|wohlfart|carle}@net.in.tum.de

## ABSTRACT

We present MoonGen, a flexible high-speed packet generator. Every so often it can saturate 10 GbE links with minimum-sized packets while using only a single CPU core by running on top of the packet processing framework DPDK. Linear multi-core scaling allows for even higher rates: We have tested MoonGen with up to 178.5 Mpps at 120 Gbit/s. Moving the whole packet generation logic into user-controlled Lua scripts allows us to achieve the highest possible flexibility. In addition, we utilize hardware features of commodity NICs that have not been used for packet generators previously. A key feature is the measurement of latency with sub-microsecond precision and accuracy by using hardware timestamping capabilities of modern commodity NICs. We address timing issues with software-based packet generators and apply methods to mitigate them with both hardware support and with a novel method to control the inter-packet gap in software. Features that were previously only possible with hardware-based solutions are now provided by MoonGen on commodity hardware. MoonGen is available as free software under the MIT license in our git repository at https://github.com/emmericp/MoonGen.

## Categories and Subject Descriptors

C.4 [**Performance of Systems**]: Measurement techniques

## 1. INTRODUCTION

Tools for traffic generation are essential to quantitative evaluations of network performance. Hardware-based solutions for packet generation are expensive and in many cases inflexible. Existing software solutions often lack performance or flexibility and come with precision problems [2].

The state of the art in packet generation, discussed further in Section 2, motivated us to design MoonGen. Our novel software packet generator is flexible, fast, and precise without relying on special-purpose hardware. Moving the packet generation logic into user-controlled Lua scripts ensures flexibility. We build on the JIT compiler LuaJIT [20] and the packet processing framework DPDK [14]. Our architecture and its implementation are described in detail in Section 3. This combination allows us to send 14.88 Mpps, line rate at 10 GbE with minimum-sized packets, from a single CPU core while executing script code for each packet. Explicit support for multi-core architectures allows us to load multiple 10 GbE interfaces simultaneously: We have tested MoonGen with 178.5 Mpps, line rate at 120 Gbit/s using twelve 2 GHz CPU cores.

MoonGen is controlled through its API instead of configuration files. We explain the interface in Section 4 by presenting code examples for typical use cases. The API allows for applications beyond packet generation as it makes DPDK packet processing facilities available to Lua scripts. Section 5 evaluates the performance of our approach. We show that running Lua code for each packet is feasible and can even be faster than an implementation written in C.

Our packet generator can also receive packets and measure round-trip latencies with sub-microsecond precision and accuracy. We achieve this by using hardware features of Intel commodity NICs that are intended for clock synchronization across networks. Section 6 features a detailed evaluation.

Section 7 investigates different methods for rate limiting on NICs. It focuses on established methods for traffic rate limiting by controlling the inter-departure times of packets based on either software mechanisms or explicit hardware features on modern NICs. Aiming for a more generic and more powerful approach to traffic limiting, Section 8 proposes a new mechanism introduced by MoonGen. This solution allows generating complex traffic patterns without additional hardware support.

MoonGen is available as free software under the MIT license [5]. Section 9 describes how to use the published code to reproduce all experiments in this paper.

## 2. STATE OF THE ART

Packet generators face a tradeoff between complexity and performance. This is reflected in the available packet generators: Barebone high-speed packet generators with limited capabilities on the one hand and feature-rich packet generators that do not scale to high data rates on the other hand. While high-speed packet generators often only send out pre-crafted Ethernet frames (e.g., pcap files), more advanced packet generators are able to transmit complex load patterns by implementing and responding to higher-layer protocols (e.g., web server load tester). Consequently, there is a lack of fast *and* flexible packet generators. Besides mere traffic generation, many packet generators also offer the possibility to capture incoming traffic and relate the generated to the received traffic.

The traditional approach to measure the performance of network devices uses hardware solutions to achieve high packet rates and high accuracy [2]. Especially their ability to accurately control the sending rate and precise timestamping are important in these scenarios. Common hardware packet generators manufactured by IXIA, Spirent, or XENA are tailored to special use cases such as performing RFC 2544 compliant device tests [3]. They send predefined traces of higher-layer protocols, but avoid complex hardware implementations of protocols. Therefore, these hardware appliances are on the fast-but-simple end of the spectrum of packet generators. They are focused on well-defined and reproducible performance tests for comparison of networking devices via synthetic traffic. However, the high costs severely limit their usage [2].

NetFPGA is an open source FPGA-based NIC that can be used as a packet generator [17]. Although costs are still beyond commodity hardware costs, it is used more often in academic publications. For example, in 2009, Covington et al. [4] described an open-source traffic generator based on NetFPGA with highly accurate inter-packet delays. The OFLOPS framework by Rotsos et al. [24] is able to measure latencies with nanosecond accuracy via a NetFPGA.

Software packet generators running on commodity hardware are widespread for different use cases. Especially traffic generators that emulate realistic traffic, e.g., Harpoon [26], suffer from poor performance on modern 10 GbE links. We focus on high-speed traffic generators that are able to saturate 10 GbE links with minimum-sized packets, i.e., achieve a rate of 14.88 Mpps. Bonelli et al. [1] implement a software traffic generator, which is able to send 12 Mpps by using multiple CPU cores. Software packet generators often rely on frameworks for efficient packet transmission [18, 23, 14] to increase the performance further to the line rate limit. Less complex packet generators can be found as example applications for high-speed packet IO frameworks: zsend for PF_RING ZC [18] and pktgen for netmap [23]. Wind River Systems provides Pktgen-DPDK [27] for DPDK [14]. Pktgen-DPDK features a Lua scripting API that can be used to control the parameters of the generator, but the scripts cannot modify the packets themselves. Further, existing tools for packet generation like Ostinato have been ported to DPDK to improve their performance [19]. Previous studies showed that software solutions are not able to precisely control the inter-packet delays [2, 4]. This leads to micro-bursts and jitter, a fact that impacts the reproducibility and validity of tests that rely on a precise definition of the generated traffic.

Ostinato is the most flexible software packet solution of the investigated options as it features configuration through Python scripts while using DPDK for high-speed packet IO. However, its scripting API is limited to the configuration of predefined settings, the scripts cannot be executed for each packet. Precise timestamping and rate control are also not supported. [19]

One has to make a choice between flexibility (software packet generators) and precision (hardware packet generators) with the available options. Today different measurement setups therefore require different packet generators. For example, precise latency measurements currently require hardware solutions. Complex packet generation (e.g., testing advanced features of network middleboxes like firewalls) requires flexible software solutions. We present a hybrid solution with the goal to be usable in all scenarios.

## 3. IMPLEMENTATION

We identified the following requirements based on our goal to close the gap between software and hardware solutions by combining the advantages of both. MoonGen must...

(R1) ...be implemented in software and run on commodity hardware.

(R2) ...be able to saturate multiple 10 GbE links with minimum-sized packets.

(R3) ...be as flexible as possible.

(R4) ...offer precise and accurate timestamping and rate control.

The following building blocks were chosen based on these requirements.

### 3.1 Packet Processing with DPDK

Network stacks of operating systems come with a high overhead [23]. We found the performance too low to fulfill requirement (R2). Packet IO frameworks like DPDK [14], PF_RING ZC [18], and netmap [23] circumvent the network stack and provide user space applications exclusive direct access to the DMA buffers to speed up packet processing. All of them have been used to implement packet generators that fulfill requirement (R2) [18, 23, 27]. We have investigated the performance of these frameworks in previous work [6] and found that DPDK and PF_RING ZC are slightly faster than netmap.

We chose DPDK for MoonGen as it supports a wide range of NICs by multiple vendors (Intel, Emulex, Mellanox, and Cisco), is well-documented, fast, and available under the BSD license [14]. PF_RING ZC was not considered further as some parts of this framework, which are needed for high-speed operation, require purchasing a license. In netmap, user space applications do not have access to the NIC's registers. This is a safety precaution as a misconfigured NIC can crash the whole system by corrupting memory [23]. This restriction in netmap is critical as it is designed to be part of an operating system: netmap is already in the FreeBSD kernel [22]. However, MoonGen needs to access NIC registers directly to implement requirement (R4).

### 3.2 Scripting with LuaJIT

MoonGen must be as flexible as possible (R3). Therefore, we move the whole packet generation logic into user-defined scripts as this ensures the maximum possible flexibility. LuaJIT was selected because related work shows that it is suitable for high-speed packet processing tasks [7] at high packet rates (R2). Its fast and simple foreign function interface allows for an easy integration of C libraries like DPDK [20].

LuaJIT may introduce unpredictable pause times due to garbage collection and compilation of code during run time.

This can lead to exhausted receive queues or starving transmission queues. Pause times introduced by the JIT compiler are in the range of "a couple of microseconds" [21]. The garbage collector (GC) works in incremental steps, the pause times depend on the usage. All packet buffers are handled by DPDK and are invisible to the GC. A typical transmit loop does not allocate new objects in Lua, so the GC can even be disabled for most experiments.

Pause times are handled by the NIC buffers: The currently supported NICs feature buffer sizes in the order of hundreds of kilobytes [11, 12, 13]. For example, the smallest buffer on the X540 chip is the 160 kB transmit buffer, which can store 128 µs of data at 10 GbE. This effectively conceals short pause times. These buffer sizes were sufficient for all of our tests.

### 3.3 Hardware Architecture

Understanding how the underlying hardware works is important for the design of a high-speed packet generator. The typical operating system socket API hides important aspects of networking hardware that are crucial for the design of low-level packet processing tools.

A central feature of modern commodity NICs is support for multi-core CPUs. Each NIC supported by DPDK features multiple receive and transmit queues per network interface. This is not visible from the socket API of the operating system as it is handled by the driver [10]. For example, both the X540 and 82599 10 GbE NICs support 128 receive and transmit queues. Such a queue is essentially a virtual interface and they can be used independently from each other. [12, 13]

Multiple transmit queues allow for perfect multi-core scaling of packet generation. Each configured queue can be assigned to a single CPU core in a multi-core packet generator. Receive queues are also statically assigned to threads and the incoming traffic is distributed via configurable filters (e.g., Intel Flow Director) or hashing on protocol headers (e.g., Receive Side Scaling). [12, 13] Commodity NICs also often support timestamping and rate control in hardware. This allows us to fulfill (R1) without violating (R4).

MoonGen does not run on arbitrary commodity hardware, we are restricted to hardware that is supported by DPDK [14] and that offers support for these features. We currently support hardware features on Intel 82599, X540, and 82580 chips. Other NICs that are supported by DPDK but not yet explicitly by MoonGen can also be used, but without hardware timestamping and rate control.

### 3.4 Software Architecture

MoonGen's core is a Lua wrapper for DPDK that provides utility functions required by a packet generator. The MoonGen API comes with functions that configure the underlying hardware features like timestamping and rate control. About 80% of the current code base is written in Lua, the remainder in C and C++. Although our current focus is on packet generation, MoonGen can also be used for arbitrary packet processing tasks.

Figure 1 shows the architecture of MoonGen. It runs a user-provided script, the *userscript*, on start-up. This script contains the main loop and the packet generation logic.

The userscript will be executed in the *master task* initially by calling the *master* function provided by the script. This master function must initialize the used NICs, i.e., config-ure the number of hardware queues, buffer sizes and filters for received traffic. It can then spawn new instances of itself running in *slave tasks* and pass arguments to them. A slave task runs a specified *slave* function. It usually receives a hardware queue as an argument and then transmits or receives packets via this queue. Starting a new slave task spawns a completely new and independent LuaJIT VM that is pinned to a CPU core. Tasks only share state through the underlying MoonGen library which offers inter-task communication facilities such as pipes. All functions related to packet transmission and reception in MoonGen and DPDK are lock-free to allow for multi-core scaling.

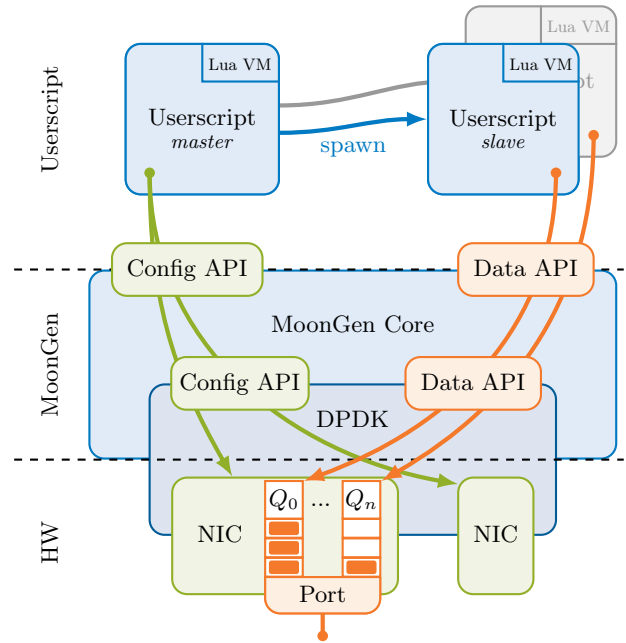

Figure 1: MoonGen's architecture

MoonGen comes with example scripts for generating load with IPv4, IPv6, IPsec, ICMP, UDP, and TCP packets, measuring latencies, measuring inter-arrival times, and generating different inter-departure times like a Poisson process and bursty traffic.

## 4. SCRIPTING API

Our example scripts in the git repository are designed to be self-explanatory exhaustive examples for the MoonGen API [5]. The listings in this section show excerpts from the `quality-of-service-test.lua` example script. This script uses two transmission tasks to generate two types of UDP flows and measures their throughput and latencies. It can be used as a starting point for a test setup to benchmark a forwarding device or middlebox that prioritizes real-time traffic over background traffic.

The example code in this section is slightly different from the example code in the repository: it has been edited for brevity. Error handling code like validation of command-line arguments is omitted here. The timestamping task has been removed as this example focuses on the basic packet generation and configuration API. Most comments have been removed and some variables renamed. The interested reader is referred to our repository [5] for the full example code including timestamping.

```
1  function master(txPort, rxPort, fgRate, bgRate)
2      local tDev = device.config(txPort, 1, 2)
3      local rDev = device.config(rxPort)
4      device.waitForLinks()
5      tDev:getTxQueue(0):setRate(bgRate)
6      tDev:getTxQueue(1):setRate(fgRate)
7      mg.launchLua("loadSlave", tDev:getTxQueue(0), 42)
8      mg.launchLua("loadSlave", tDev:getTxQueue(1), 43)
9      mg.launchLua("counterSlave", rDev:getRxQueue(0))
10     mg.waitForSlaves()
11 end
```

Listing 1: Initialization and device configuration

## 4.1 Initialization

Listing 1 shows the `master` function. This function is executed in the master task on startup and receives command line arguments passed to MoonGen: The devices and transmission rates to use in this case. It configures one transmit device with two transmit queues and one receiving device with the default settings. The call in line 4 waits until the links on all configured devices are established. It then configures hardware rate control features on the transmission queues and starts three slave threads, the first two generate traffic, the last counts the received traffic on the given device. The arguments passed to `mg.launchLua` are passed to the respective functions in the new task. The `loadSlave` function takes the transmission queue to operate on and a port to distinguish background from prioritized traffic.

## 4.2 Packet Generation Loop

Listing 2 shows the `loadSlave` function that is started twice and does the actual packet generation. It first allocates a memory pool, a DPDK data structure in which packet buffers are allocated. The MoonGen wrapper for memory pools expects a callback function that is called to initialize each packet. This allows a script to fill all packets with default values (lines 5 to 10) before the packets are used in the transmit loop (lines 17 to 24). The transmit loop only needs to modify a single field in each transmitted packet (line 20) to generate packets from randomized IP addresses.

Line 13 initializes a packet counter that keeps track of transmission statistics and prints them in regular intervals. MoonGen offers several types of such counters with different methods to acquire statistics, e.g., by reading the NICs statistics registers. This example uses the simplest type, one that must be manually updated.

Line 15 allocates a `bufArray`, a thin wrapper around a C array containing packet buffers. This is used instead of a normal Lua array for performance reasons. It contains a number of packets in order to process packets in batches instead of passing them one-by-one to the DPDK API. Batch processing is an important technique for high-speed packet processing [6, 23].

The main loop starts in line 17 with allocating packets of a specified size from the memory pool and storing them in the packet array. It loops over the newly allocated buffers (line 18) and randomizes the source IP (line 20). Finally, checksum offloading is enabled (line 22) and the packets are transmitted (line 23).

Note that the main loop differs from a packet generator relying on a classic API. MoonGen, or any other packet generator based on a similar framework, cannot simply re-use buffers because the transmit function is asynchronous. Passing packets to the transmit function merely places pointers

```
1  local PKT_SIZE = 124
2  function loadSlave(queue, port)
3      local mem = memory.createMemPool(function(buf)
4          buf:getUdpPacket():fill{
5              pktLength = PKT_SIZE,
6              ethSrc = queue, -- get MAC from device
7              ethDst = "10:11:12:13:14:15",
8              ipDst = "192.168.1.1",
9              udpSrc = 1234,
10             udpDst = port,
11         }
12     end)
13     local txCtr = stats:newManualTxCounter(port, "plain")
14     local baseIP = parseIPAddress("10.0.0.1")
15     local bufs = mem:bufArray()
16     while dpdk.running() do
17         bufs:alloc(PKT_SIZE)
18         for _, buf in ipairs(bufs) do
19             local pkt = buf:getUdpPacket()
20             pkt.ip.src:set(baseIP + math.random(255) - 1)
21         end
22         bufs:offloadUdpChecksums()
23         local sent = queue:send(bufs)
24         txCtr:updateWithSize(sent, PKT_SIZE)
25     end
26     txCtr:finalize()
27 end
```

Listing 2: Transmission slave task

to them into a memory queue, which is accessed by the NIC later [14]. A buffer must not be modified after passing it to DPDK. Otherwise, the transmitted packet data may be altered if the packet was not yet fetched by the NIC.

Therefore, we have to allocate new packet buffers from the memory pool in each iteration. Pre-filling the buffers at the beginning allows us to only touch fields that change per packet in the transmit loop. Packet buffers are recycled by DPDK in the transmit function, which collects packets that were sent by the NIC earlier [14]. This does not erase the packets' contents.

## 4.3 Packet Counter

Listing 3 shows how to use MoonGen's packet reception API to measure the throughput of the different flows by counting the incoming packets.

The task receives packets from the provided queue in the `bufArray bufs` in line 5. It then extracts the UDP destination port from the packet (line 8) and uses counters to track statistics per port. The final statistics are printed by calling the counters' finalize methods in line 19. Printed statistics

```
1  function counterSlave(queue)
2      local bufs = memory.bufArray()
3      local counters = {}
4      while dpdk.running() do
5          local rx = queue:recv(bufs)
6          for i = 1, rx do
7              local buf = bufs[i]
8              local port = buf:getUdpPacket().udp:getDstPort()
9              local ctr = counters[port]
10             if not ctr then
11                 ctr = stats:newPktRxCounter(port, "plain")
12                 counters[port] = ctr
13             end
14             ctr:countPacket(buf)
15         end
16         bufs:freeAll()
17     end
18     for _, ctr in pairs(counters) do
19         ctr:finalize()
20     end
21 end
```

Listing 3: Packet counter slave task

include the average packet and byte rates as well as their standard deviations.

The format to print in is specified in the counter constructor in line 11. All example scripts use the `plain` formatter, the default value is `CSV` for easy post-processing. The output can also be diverted to a file. Details are in the documentation of `stats.lua`.

This script can be used for another similar test setup by adapting the code to the test setup by changing hardcoded constants like the used addresses and ports. The full script in the repository [5] includes a separate timestamping task to acquire and print latency statistics for the two flows.

# 5. PERFORMANCE

Writing the whole generation logic in a scripting language raises concerns about the performance. One important feature of LuaJIT is that it allows for easy integration with existing C libraries and structs: it can directly operate on C structs and arrays without incurring overhead for bound checks or validating pointers [20]. Thus, crafting packets is very efficient in MoonGen.

The obvious disadvantage is that unchecked memory accesses can lead to memory corruption, a problem that is usually absent from scripting languages. However, most critical low-level parts like the implementation of the NIC driver are handled by DPDK. The MoonGen core then wraps potentially unsafe parts for the userscript if possible. There are only two operations in a typical userscript that can lead to memory corruption: writing beyond packet buffer boundaries and trying to operate on buffers that are null pointers. This is an intentional design decision to aid the performance by avoiding unnecessary checks.

## 5.1 Test Methodology

Measuring the CPU load caused by a DPDK-based application is difficult because DPDK recommends a busy-wait loop [14], i.e., the CPU load is always 100% for a typical application. MoonGen and other DPDK-based generators like Pktgen-DPDK [27] are no exceptions to this. The bottleneck for packet transmission is usually not the CPU but the line rate of the network, so just measuring the achieved rate provides no insight. We therefore decrease the clock frequency of our CPU such that the processing power becomes the bottleneck. The performance can then be quantified as CPU cycles per packet. The same approach was used by Rizzo to evaluate the performance of netmap [23].

The tests in this section were executed on an Intel Xeon E5-2620 v3 CPU with a frequency of 2.4 GHz that can be clocked down to 1.2 GHz in 100 MHz steps. To ensure consistent and reproducible measurement results, we disabled Hyper-Threading, which may influence results if the load of two virtual cores is scheduled to the same physical core. TurboBoost and SpeedStep were also disabled because they adjust the clock speed according to the current CPU load and interfere with our manual adjustment of the frequency.

## 5.2 Comparison with Pktgen-DPDK

Our scripting approach can even increase the performance compared to a static packet generator slightly. We show this by comparing MoonGen to Pktgen-DPDK 2.5.1 [27], a packet generator for DPDK written in C.

We configured both packet generators to craft minimum-sized UDP packets with 256 varying source IP addresses on a single CPU core. We then gradually increased the CPU's frequency until the software achieved line rate. Pktgen-DPDK required 1.7 GHz to hit the 10 GbE line rate of 14.88 Mpps, MoonGen only 1.5 GHz. Pktgen-DPDK achieved 14.12 Mpps at 1.5 GHz. This means MoonGen is more efficient in this specific scenario.

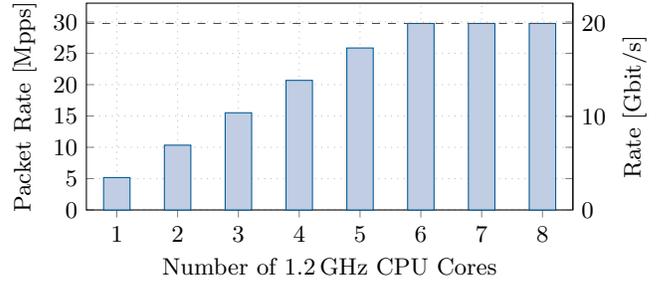

Figure 2: Multi-core scaling under high load

This increased performance is an inherent advantage of MoonGen's architecture: Pktgen-DPDK needs a complex main loop that covers all possible configurations even though we are only interested in changing IP addresses in this test scenario. MoonGen, on the other hand, can use a script that consists of a tight inner loop that exclusively executes the required tasks: allocating pre-filled packet buffers, modifying the IP address, and sending the packets with checksum offloading. You only pay for the features you actually use with MoonGen.

## 5.3 Multi-core Scaling

The achieved performance depends on the script; the previous example was a light workload for the comparison to Pktgen-DPDK, which is limited to such simple patterns. Therefore, we test a more involved script to stress MoonGen to show the scaling with multiple cores sending to the same NIC via multiple transmission queues.

Figure 2 shows the performance under heavy load and the scaling with the number of CPU cores. MoonGen was configured to generate minimum-sized packets with random payload as well as random source and destination addresses and ports. The code generates 8 random numbers per packet to achieve this. Each core generated and sent packets on two different 10 GbE interfaces simultaneously. Linear scaling can be observed up to the line rate limit (dashed line).

The code was written in idiomatic Lua without specific optimizations for this use case: LuaJIT's standard random number generator, a Tausworthe generator [20], was used. Since a high quality random number generator is not required here, a simple linear congruential generator would be faster. The code also generates a random number per header field instead of combining multiple fields (e.g., source and destination port can be randomized by a single 32-bit random number).

Despite the lack of optimizations, the code was initially found to be too fast for meaningful scalability measurements (10.3 Mpps on a single core). We therefore reduced the CPU's clock speed to 1.2 GHz and increased the number of NICs to 2 for this test. This test shows that sending to a single NIC port via multiple queues scales linearly, an important assumption made for our architecture (cf. Section 3.3).

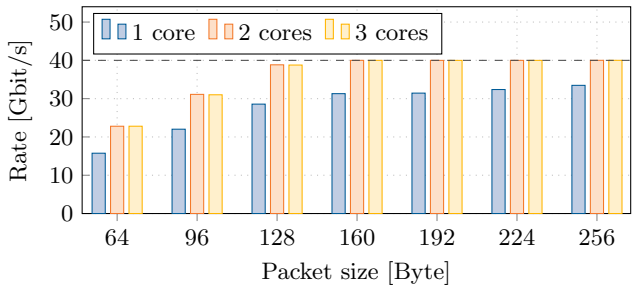

**Figure 3: Throughput with an XL710 40 GbE NIC**

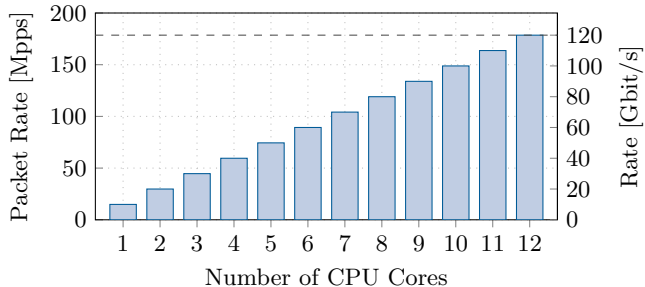

**Figure 4: Multi-core scaling (multiple 10 GbE NICs)**

### 5.4 Scaling to 40 Gigabit Ethernet

40 GbE NICs like the dual port Intel XL710 [15] are currently being introduced to the market. However, these first generation NICs come with bandwidth limitations that do not exist on the 10 GbE NICs discussed previously: they cannot saturate a link with minimum-sized packets [16] and they cannot saturate both ports simultaneously regardless of the packet size [15]. This may limit their use in some scenarios where a large number of small packets is required, e.g., stress-testing a router.

We are currently adding support for these NICs in MoonGen and present preliminary results here. Figure 3 shows the achieved throughput with various packet sizes and number of 2.4 GHz CPU cores used to generate the traffic. Packet sizes of 128 bytes or less cannot be generated in line rate. Using more than two CPU cores does not improve the speed, so this is a hardware bottleneck as described by Intel [16].

The second bandwidth restriction of this NIC is the aggregate bandwidth of the two ports. One obvious restriction is the 63 Gbit/s bandwidth of the PCIe 3.0 x8 link that connects the NIC to the CPU. However, the main bottleneck is the media access control layer in the XL710 chip: it is limited to a maximum aggregate bandwidth of 40 Gbit/s (cf. Section 3.2.1 of the XL710 datasheet [15]). We could achieve a maximum bandwidth of 50 Gbit/s with large packets on both ports simultaneously and a maximum packet rate of 42 Mpps (28 Gbit/s with 64 byte frames).

### 5.5 Scaling to 100 Gigabit Ethernet

100 GbE is currently restricted to hardware appliances like switches and routers and not yet available on commodity server hardware. We can emulate higher speeds by using multiple NICs.

We equipped one of our test servers with six dual-port 10 GbE Intel X540-T2 NICs to investigate the performance at high rates. Figure 4 shows the achieved packet rate when generating UDP packets from varying IP addresses. We used two Intel Xeon E5-2640 v2 CPUs with a nominal clock rate of 2 GHz for this test, but the clock rate can even be reduced to 1.5 GHz for this packet generation task (cf. Section 5.2).

Note that sending to multiple NICs simultaneously is architecturally the same as sending to multiple queues on a single NIC as different queues on a single NIC are independent from each other (cf. Section 5.3) in an ideal well-behaved NIC like the current generation of 10 GbE NICs. We do not expect significant challenges when moving to 100 GbE due to this architecture and promising tests with multiple 10 GbE ports. However, the first generation of 100 GbE NICs will likely have similar hardware restrictions as the 40 GbE NICs discussed in Section 5.4 which need to be taken into account.

| Operation | Cycles/Pkt |
|---|---|
| Packet transmission | $76.0 \pm 0.8$ |
| Packet modification | $9.1 \pm 1.2$ |
| Packet modification (two cachelines) | $15.0 \pm 1.3$ |
| IP checksum offloading | $15.2 \pm 1.2$ |
| UDP checksum offloading | $33.1 \pm 3.5$ |
| TCP checksum offloading | $34.0 \pm 3.3$ |

**Table 1: Per-packet costs of basic operations**

### 5.6 Per-Packet Costs

MoonGen's dynamic approach to packet generation in userscripts does not allow for a performance analysis in a general configuration as there is no typical scenario. Nevertheless, the cost of sending a packet can be decomposed into three main components: packet IO, memory accesses, and packet modification logic. We devised a synthetic benchmark that measures the average number of CPU cycles required for various operations that are commonly found in packet generator scripts. These measurements can be used to estimate the hardware requirements of arbitrary packet generator scripts. We repeated all measurements ten times; the uncertainties given in this section are the standard deviations.

#### 5.6.1 Basic Operations

Table 1 shows the average per-packet costs of basic operations for IO and memory accesses. The baseline for packet IO consists of allocating a batch of packets and sending them without touching their contents in the main loop. This shows that there is a considerable per-packet cost for the IO operation caused by the underlying DPDK framework.

Modification operations write constants into the packets, forcing the CPU to load them into the layer 1 cache. Additional accesses within the same cache line (64 bytes) add no measurable additional cost. Accessing another cache line in a larger packet is noticeable.

Offloading checksums is not free (but still cheaper than calculating them in software) because the driver needs to set several bitfields in the DMA descriptor. For UDP and TCP offloading, MoonGen also needs to calculate the IP pseudo header checksum as this is not supported by the X540 NIC used here [13].

#### 5.6.2 Randomizing Packets

Sending varying packets is important to generate different flows. There are two ways to achieve this: one can either generate a random number per packet or use a counter with wrapping arithmetic that is incremented for each packet. The resulting value is then written into a header field. Table 2 shows the cost for the two approaches, the baseline is the cost of writing a constant to a packet and sending it (85.1 cycles/pkt).

| Fields | Cycles/Pkt (Rand) | Cycles/Pkt (Counter) |
|--------|-------------------|----------------------|
| 1      | $32.3 \pm 0.5$    | $27.1 \pm 1.4$       |
| 2      | $39.8 \pm 1.0$    | $33.1 \pm 1.3$       |
| 4      | $66.0 \pm 0.9$    | $38.1 \pm 2.0$       |
| 8      | $133.5 \pm 0.7$   | $41.7 \pm 1.2$       |

Table 2: Per-packet costs of modifications

There is a fixed cost for calculating the values while the marginal cost is relatively low: 17 cycles/pkt per random field and 1 cycle/pkt for wrapping counters. These results show that wrapping counters instead of actual random number generation should be preferred if possible for the desired traffic scenario.

### 5.6.3 Cost Estimation Example

We can use these values to predict the performance of the scripts used for the performance evaluation in Section 5.3. The example generated 8 random numbers for fields with a userscript that is completely different from the benchmarking script: it writes the values into the appropriate header fields and the payloads, the benchmarking script just fills the raw packet from the start. The script also combines offloading and modification; the benchmark tests them in separate test runs.

The expected cost consists of: packet IO, packet modification, random number generation, and IP checksum offloading, i.e., $229.2 \pm 3.9$ cycles/pkt. This translates to a predicted throughput of $10.47 \pm 0.18$ Mpps on a single 2.4 GHz CPU core. The measured throughput of 10.3 Mpps is within that range. This shows that our synthetic benchmark can be used to estimate hardware requirements.

## 5.7 Effects of Packet Sizes

All tests performed in the previous sections use minimum-sized packets. The reason for this choice is that the per-packet costs dominate over costs incurred by large packets. Allocating and sending larger packets without modifications add no additional cost in MoonGen on 1 and 10 GbE NICs. Only modifying the content on a per-packet basis adds a performance penalty, which is comparatively low compared to the fixed cost of sending a packet. Using larger packets also means that fewer packets have to be sent at line rate, so the overall fixed costs for packet IO are reduced: minimum-sized packets are usually the worst-case.

Nevertheless, there are certain packet sizes that are of interest: those that are just slightly larger than a single cache line. We benchmarked all packet sizes between 64 and 128 bytes and found no difference in the CPU cycles required for sending a packet. Since MoonGen also features packet reception, we also tried to receive packets with these sizes and found no measurable impact of the packet size.[1]

Rizzo notes that such packet sizes have a measurable impact on packet reception, but not transmission, in his evaluation of netmap [23]. He attributes this to hardware bottlenecks as it was independent from the CPU speed. We could not reproduce this with MoonGen. The likely explanation is that we are using current (2014) server hardware, while the evaluation of netmap was done in 2012 on an older system with a CPU launched in 2009 [23].

## 6. HARDWARE TIMESTAMPING

Another important performance characteristic beside the throughput is the latency of a system. Modern NICs offer hardware support for the IEEE 1588 Precision Time Protocol (PTP) for clock synchronization across networks. PTP can be used either directly on top of Ethernet as a layer 3 protocol with EtherType 0x88F7 or as an application-layer protocol on top of UDP [8].

We examined the PTP capabilities of the Intel 82580 GbE and the 82599 and X540 10 GbE chips. They support timestamping of PTP Ethernet and UDP packets, the UDP port is configurable on the 10 GbE NICs. They can be configured to timestamp only certain types of PTP packets, identified by the first byte of their payload. The second byte must be set to the PTP version number. All other PTP fields in the packet are not required to enable timestamps and may contain arbitrary values. [11, 12, 13] This allows us to measure latencies of almost any type of packet.

Most Intel NICs, including all 10 GbE chips, save the timestamps for received and transmitted packets in a register on the NIC. This register must be read back before a new packet can be timestamped [12, 13], limiting the throughput of timestamped packets. Some Intel GbE chips like the 82580 support timestamping all received packets by prepending the timestamp to the packet buffer [11].

### 6.1 Precision and Accuracy

Timestamping mechanisms of the Intel 82599 and Intel X540 10 GbE chips operate at 156.25 MHz when running at 10 GbE speeds [12, 13]. This frequency is reduced to 15.625 MHz when a 1 GbE link is used, resulting in a precision of 6.4 ns for 10 GbE and 64 ns for 1 GbE. The datasheet of the Intel 82580 GbE [11] controller lacks information about the clock frequency. Testing shows that the acquired timestamps are always of the form $t = n \cdot 64\,ns + k \cdot 8\,ns$ where $k$ is a constant that varies between resets, so the precision is 64 ns.

All of these NICs timestamp packets late in the transmit path and early in the receive path to be as accurate as possible [11, 12, 13]. We tested the timestamping functionality by using loop-back multimode OM3 fiber cables on an 82599-based NIC with a 10GBASE-SR SFP+ module and Cat 5e cable between the two ports of a dual-port X540-based NIC. Table 3 on the next page shows measured latencies $t_x$ for different cable lengths $x$ for each NIC as well as the (de-)modulation time $k$ and propagation speed $v_p$, which can be calculated from these data points with the equation $t = k + l/v_p$. $k$ is higher on the copper-based NIC, this is likely due to the more complex line code required for 10GBASE-T [9]. This calculation does not take any errors in the cable length into account; we rely on the vendor's specification[2]. The actual propagation speed and encoding times may therefore be outside the interval given in Table 3.

We repeated each measurement at least 500 000 times. All measurements for the fiber connection on the 82599 NIC

---

[1] Note that this is not true for XL710 40 GbE NICs which can run into hardware bottlenecks with some packet sizes.

[2] We believe that the 50 m cable is actually slightly shorter.

| NIC | $t_{2m}$ [ns] | $t_{8.5m}$ [ns] | $t_{10m}$ [ns] | $t_{20m}$ [ns] | $t_{50m}$ [ns] | $k$ [ns] | $v_p$ |
|---|---|---|---|---|---|---|---|
| 82599 (fiber) | 320 | 352 | - | 403.2 | - | $310.7 \pm 3.9$ | $0.72c \pm 0.056c$ |
| X540 (copper) | 2156.8 | - | 2195.2 | - | 2387.2 | $2147.2 \pm 4.8$ | $0.69c \pm 0.019c$ |

Table 3: Timestamping accuracy measurements

yielded the same result except for the 8.5 m cable. This cable caused a latency of 345.6 ns in 50.2% of the measurements and 358.4 ns in the other 49.8% (Table 3 shows the average). This variance is due to the fact that the timer that is saved when the timestamp is taken is incremented only every two clock cycles on the 82599 chip [12], i.e., the granularity of the timer is 12.8 ns but the timestamping operates at 6.4 ns.

The timestamp timer on the X540 is incremented every 6.4 ns so it does not have this problem. However, it faces a different challenge: the 10GBASE-T standard uses a block code on layer 1 [9] which introduces variance. Table 3 shows the median latency. More than 99.5% of the measured values were within $\pm$ 6.4 ns of the median. The difference between the minimum and maximum latencies was 64 ns. These variations were independent of the cable length.

The absence of a variance on the 82599 chip demonstrates a high precision, the plausible results for the modulation time [28] and the linear behavior of the propagation latency show a high accuracy.

## 6.2 Clock Synchronization

Test setups can involve multiple network ports that may even be on different NICs. For example, measuring the forwarding latency of a switch requires timestamping a packet on two different ports. MoonGen therefore needs to be able to synchronize the clocks between two network ports. This is even necessary between two ports of a dual-port NIC, which are completely independent from the user's point of view.

MoonGen synchronizes the clocks of two ports by reading the current time from both clocks and calculating the difference. The clocks are then read again in the opposite order. The resulting differences are the same if and only if the clocks are currently synchronous (assuming that the time required for the PCIe access is constant). We observed randomly distributed outliers in about 5% of the reads. We therefore repeat the measurement 7 times to have a probability of > 99.999% of at least 3 correct measurements. The median of the measured differences is then used to adjust one of the clocks to synchronize them. This adjustment must be done with an atomic read-modify-write operation. The NICs support this as it is also required for PTP.

Tests show that this technique synchronizes the clocks with an error of ±1 cycle. Therefore, the maximum accuracy for tests involving multiple network interfaces is 19.2 ns for the 10 GbE chips.

## 6.3 Clock Drift

Using two different clocks also entails the risk of clock drifts. Drift on X540-based NICs depends on the physical wiring as the timestamping clock is synchronized to the physical layer. Two ports on different X540-based NICs that are directly connected do not exhibit any clock drift while the link is established. However, the clocks of two ports on the same X540 NIC will drift if they are connected to two different NICs. We measured the drift between different X540 and 82599 NICs. The worst-case observed drift was 35 µs per second between a NIC on the mainboard and a discrete NIC.

MoonGen handles clock drift by resynchronizing the clocks before a timestamped packet is sent, so this drift translates to a relative error of only 0.0035%. This is not significant for latency measurements. Since the measurements show a constant clock drift, it would also be possible to subtract the accumulated drift from the acquired timestamps to avoid resynchronization.

## 6.4 Limitations

Our approach for latency measurements comes with limitations. The latency measurements are restricted to Ethernet frames with the PTP EtherType and UDP packets. MoonGen cannot measure latencies of other protocols.

The naïve handling of clock drift by resynchronizing the clocks for each packet allows for only a single timestamped packet in flight, limiting the throughput to $1\,Pkt/RTT$. MoonGen scripts therefore use two transmission queues, one that sends timestamped packets and one that sends regular packets. The regular packets can be crafted such that the device under test cannot distinguish them from the timestamped packets, e.g., by setting the PTP type in the payload to a value that is not timestamped by the NIC. So MoonGen effectively samples random packets in the data stream and timestamps them. Note that the benchmarking standard RFC 2544 calls for only one timestamped packet in a 120 second interval [3]. MoonGen can timestamp several thousands of packets per second to calculate average latencies and histograms.

The investigated NICs refuse to timestamp UDP PTP packets that are smaller than the expected packet size of 80 bytes. Larger packets are timestamped properly. This restriction does not apply to packets with the PTP EtherType as the minimum PTP packet size is below 64 bytes in this configuration. Measurements of inter-arrival times are restricted to GbE networks due to lack of hardware support for timestamping in line rate on 10 GbE NICs.

Based on the discussed measurement results and despite these limitations, we argue that special-purpose hardware is not necessary to conduct highly precise and accurate latency and inter-arrival time measurements.

## 7. RATE CONTROL

An important feature of a packet generator is controlling the packet rate and generating specific timing patterns to simulate real-world scenarios. MoonGen utilizes hardware rate control features of Intel NICs to generate constant bit rate and bursty traffic. We also implement a novel software-based rate control for realistic traffic patterns, e.g., based on a Poisson process. That is discussed further in Section 8, this section focuses on software rate control in other packet generators and hardware rate control.

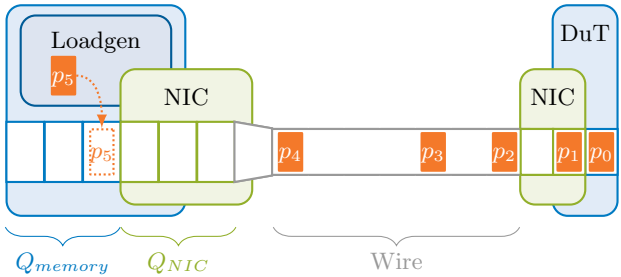

Figure 5: Software-based rate control

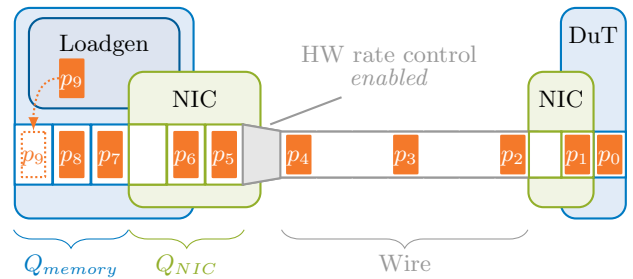

Figure 6: Hardware-based rate control

## 7.1 Software Rate Control in Existing Packet Generators

Trying to control the timing between packets in software is known to be error-prone [2, 4]. The main problem with software-based rate control is that the software needs to push individual packets to the NIC and then has to wait for the NIC to transmit it before pushing the next packet.

However, modern NICs do not work that way: they rely on an asynchronous push-pull model and not on a pure push model. The software writes the packets into a queue that resides in the main memory and informs the NIC that new packets are available. It is up to the NIC to fetch the packets asynchronously via DMA and store them in the internal transmit queue on the NIC before transmitting them. A good explanation of this packet flow can be found in the datasheet of the X540 chip [13] (Section 1.7), other NICs follow a similar process.

Figure 5 visualizes this packet flow. Only a single packet at a time is allowed in the queues ($Q_{memory}$ & $Q_{NIC}$) to generate packets that are not back-to-back on the wire.

This hardware architecture causes two problems: the exact timing when a packet is retrieved from memory cannot be controlled by the software and queues cannot be used (unless bursts are desired). The former results in a low precision, as the exact time when the packet is transferred cannot be determined. The latter impacts the performance at high packet rates as high-speed packet processing relies on batch processing [6, 23].

## 7.2 Hardware Rate Control

Intel 10 GbE NICs feature hardware rate control: all transmit queues can be configured to a specified rate. The NIC then generates constant bit-rate (CBR) traffic. This solves the two problems identified in the previous section. The software can keep all available queues completely filled and the generated timing is up to the NIC. Figure 6 shows this architecture. The disadvantage is that this approach is limited to CBR traffic and bursty traffic (by changing the rate parameter periodically).

## 7.3 Evaluation

We compare our hardware-assisted solution to the software-based rate control found in zsend 6.0.2 (an example application of the PF_RING ZC framework [18]), and Pktgen-DPDK 2.5.1 [27] to quantify the adverse effects of software-based rate control. We use an Intel 82580 GbE controller, which is able to timestamp arbitrary received packets in line rate (cf. Section 6) to measure inter-arrival times.

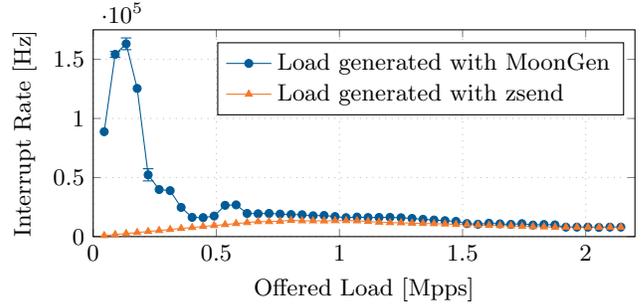

Figure 7: Interrupt rate with micro-bursts

Figure 8 on the next page compares the inter-arrival times generated by different packet generators. The generators use an X540 NIC, which also supports 1 Gbit/s. The histograms have a bin size of 64 ns (precision of the 82580 chip) and were generated by observing at least 1 000 000 packets.

Traffic from the hardware rate-controlled NIC oscillates around the targeted inter-arrival time by up to 256 ns and it avoids generating bursts almost completely (inter-arrival time of 672 ns, marked with a black arrow in Figure 8). Table 4 summarizes the results. The best result in each column is highlighted.

We discussed these findings with the authors of zsend as we configured it explicitly to avoid bursts. We then tested several suggested configurations and versions. We concluded that these observations indicate a bug in the PF_RING ZC framework that is being investigated.

It stands to reason that the precision problems as well as the micro-bursts intensify further at rates beyond 1 Gbit/s with software-based rate control. Measuring inter-arrival times on 10 GbE is a challenging task: Reliable measurements require special-purpose hardware. We do not yet have such hardware. We predict that the precision of our hardware-assisted approach will improve at 10 GbE speeds: The frequency of the internal clock on the NIC that controls the inter-departure times is scaled up by a factor of 10 when operating at 10 GbE compared to GbE [13].

## 7.4 Effects of Micro-Bursts on Linux Systems

Figure 7 visualizes the interrupt rate on a Linux packet forwarder running Open vSwitch under increasing load generated by MoonGen and zsend. Open vSwitch was configured with a static OpenFlow rule to forward between two ports. The micro-bursts generate a low interrupt rate. The likely explanation for this is that the bursts trigger the interrupt rate moderation feature of the driver earlier than expected. This shows that bad rate control can have a measurable impact on the behavior of the tested system.

| Rate | Packet Generator | Micro-Bursts | ±64 ns | ±128 ns | ±256 ns | ±512 ns |
|---|---|---|---|---|---|---|
| 500 kpps | MoonGen | 0.02% | **49.9%** | **74.9%** | **99.8%** | **99.8%** |
| | Pktgen-DPDK | **0.01%** | 37.7% | 72.3% | 92% | 94.5% |
| | zsend | 28.6% | 3.9% | 5.4% | 6.4% | 13.8% |
| 1000 kpps | MoonGen | **1.2%** | **50.5%** | 52% | **97%** | **100%** |
| | Pktgen-DPDK | 14.2% | 36.7% | **58%** | 70.6% | 95.9% |
| | zsend | 52% | 4.6% | 7.9% | 24.2% | 88.1% |

Table 4: Rate control measurements

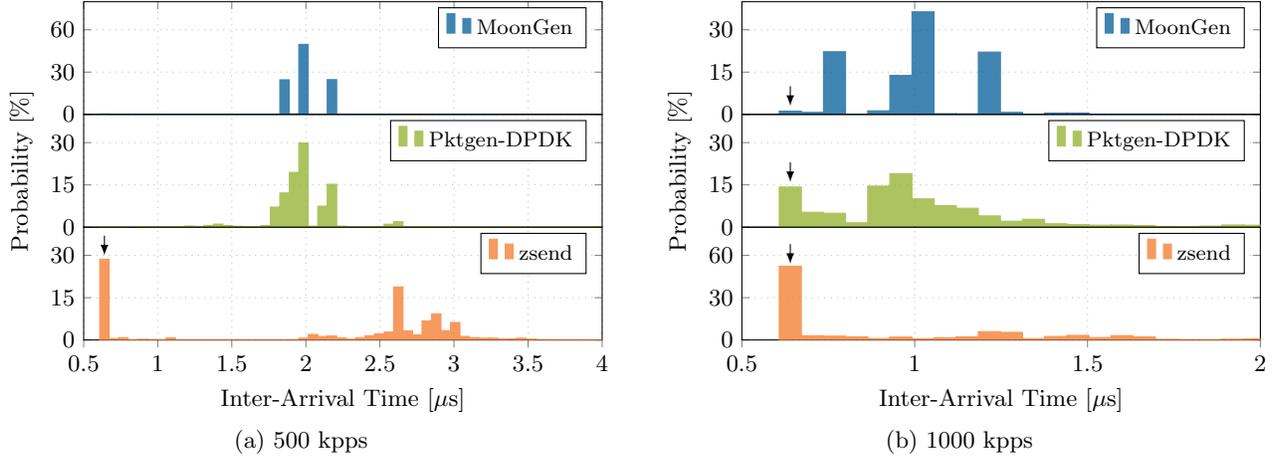

(a) 500 kpps  (b) 1000 kpps

Figure 8: Histograms of inter-arrival times

## 7.5 Limitations of Hardware Rate Control

In our tests we encountered unpredictable non-linear behavior with packet rates above ~9 Mpps (~6 Gbit/s wire-rate with 64 byte packets) on Intel X520 and X540 NICs. A work-around is configuring two transmission queues and sending a CBR stream from both of them. Note that this is not equivalent to a single transmission queue with proper rate control as both queues control their transmission rate independently from each other.

Hardware rate control of the investigated NICs is also restricted to CBR traffic, so MoonGen still needs an implementation of software-based rate control for other traffic patterns.

## 8. CONTROLLING INTER-PACKET GAPS IN SOFTWARE

To overcome this restriction to constant bit rate or bursty traffic, MoonGen implements a novel mechanism for software-based rate control. This allows MoonGen to create arbitrary traffic patterns.

### 8.1 Sending Gaps on the Wire

We were not satisfied with the precision of existing software rate control mechanisms (cf. Section 7.3 and [2, 4]) so we present a new method here. All existing packet generators try to delay sending packets by not sending packets for a specified time, leading to the previously mentioned problems. MoonGen fills the gaps between packets with invalid packets instead. Varying the length of the invalid packet precisely determines the time between any two packets and subsequently allows the creation of arbitrary complex traffic patterns. With this technique, we can still make use of the NIC's queues and do not have to rely on any timing related to DMA accesses by the NIC.

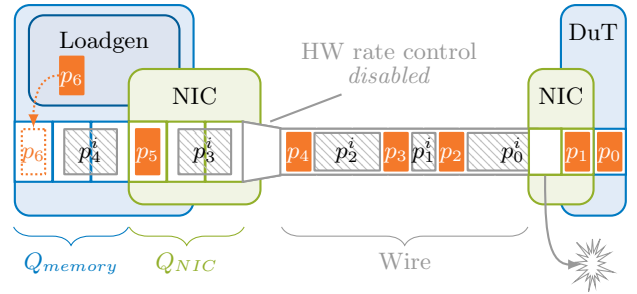

Figure 9: Precise generation of arbitrary traffic patterns in MoonGen

This approach requires support by the device under test (DuT): it needs to detect and ignore invalid packets in hardware without affecting the packet processing logic. MoonGen uses packets with an incorrect CRC checksum and, if necessary, an illegal length for short gaps. All investigated NICs in our testbed drop such packets early in the receive flow: they are dropped even before they are assigned to a receive queue, the NIC only increments an error counter [11, 12, 13]. Subsequently, the packet processing logic is not affected by this software rate control mechanism.

Figure 9 illustrates this concept. Shaded packets $p_j^i$ are sent with an incorrect CRC checksum, all other packets $p_k$ with a correct one. Note that the wire and all transmission queues are completely filled, i.e., the generated rate has to be the line rate.

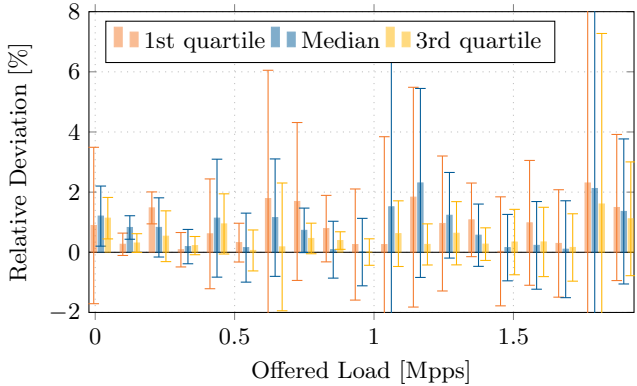

Figure 10: Differences in forwarding latencies of Open vSwitch with CBR traffic generated by hardware and our software approach

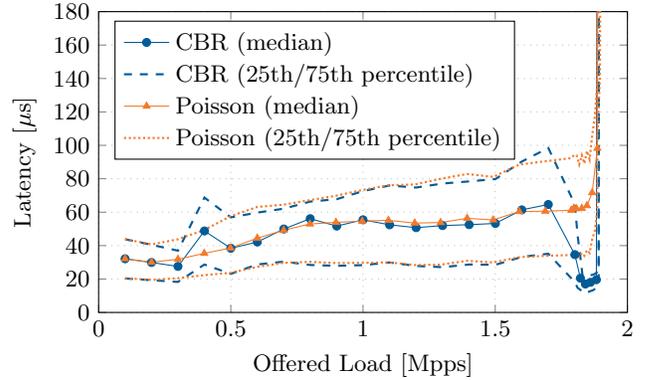

Figure 11: Forwarding latency of Open vSwitch with CBR and Poisson traffic patterns

In theory, arbitrary inter-packet gaps should be possible. The NICs we tested refused to send out frames with a wire-length (including Ethernet preamble, start-of-frame delimiter, and inter-frame gap) of less than 33 bytes, so gaps between 1 and 32 bytes (0.8 ns to 25.6 ns) cannot be generated. Generating small frames also puts the NIC under an unusually high load for which it was not designed. We found that the maximum achievable packet rate with short frames is 15.6 Mpps on Intel X540 and 82599 chips, only 5% above the line rate for packets with the regular minimal size. MoonGen therefore enforces a minimum wire-length of 76 bytes (8 bytes less than the regular minimum) by default for invalid packets. As a result, gaps between 0.8 ns and 60.8 ns cannot be represented.

### 8.2 Evaluation

We generate CBR traffic with our approach and compare it to CBR traffic generated by the hardware facilities of our NIC by comparing the response of a DuT.

We use Intel's hardware implementation as reference generator. The same measurement with other software-based packet generators is not possible as they don't support accurate timestamping. However, the results from Section 7.4 indicate that the latency would be affected at low rates due to the measurably different interrupt rate (cf. Figure 7).

Figure 10 shows the difference of the 25th, 50th, and 75th percentiles of the forwarding latency on a server running Open vSwitch. The test is restricted to the range 0.1 Mpps to 1.9 Mpps as the DuT becomes overloaded at higher rates and the latency is a function of the buffer size of the DuT after this point. We repeated the whole test 10 times, the error bars in the figure show the resulting standard deviations. The relative deviation is within $1.2\sigma$ of 0% for almost all measurement points, only the 1st quartile at 0.23 Mpps deviates by $1.5\% \pm 0.5\%$. Minor activity on the DuT, e.g., an active SSH session, shows a significantly larger ($\geq 10\%$) effect on the latency with both rate control methods. This shows that loading the DuT with a large number of invalid packets does not cause system activity; the DuT does not notice the invalid packets.

### 8.3 Example: Poisson Traffic

CBR traffic is often an unrealistic test scenario for measurements of latency. Bursts or a Poisson process allow for more sophisticated tests that also stress buffers as the DuT becomes temporarily overloaded.

Figure 11 shows measured latencies of Open vSwitch configured to forward packets between two ports. We generate packets with CBR (hardware rate control) and Poisson (CRC-based software rate control) traffic patterns and compare their latencies. The outlier at 0.4 Mpps for CBR traffic was reproducible across multiple re-measurements on different servers. The sudden drop in latency before the system becomes overloaded was also reproducible. Both are likely artifacts of the interaction between the interrupt throttle algorithm found in the Intel driver [10] and the dynamic interrupt adaption of Linux [25] on the DuT. The artifacts are present regardless of how the CBR traffic is generated (cf. Figure 10), so this is not caused by MoonGen but an effect of the traffic pattern on DuT.

The system becomes overloaded at about 1.9 Mpps, resulting in packet drops and a very large latency (about 2 ms in this test setup) as all buffers are filled. The overall achieved throughput is the same regardless of the traffic pattern and method to generate it. This result shows that the traffic pattern can affect the DuT in an experiment measurably, underlining the importance of a reliable precise packet generator.

### 8.4 Limitations of our Approach

A shorter per-byte transmission time improves both the granularity and the minimum length that can be generated. This means our solution works best on 10 GbE where the granularity is 0.8 ns.

Due to the artificially enforced minimum size of 76 bytes, gaps between 1 and 75 bytes (0.8 ns to 60 ns) cannot be precisely represented (cf. Section 8.1). It is possible to reduce this range for tests with larger packets or lower rates. We approximate gaps that cannot be generated by occasionally skipping an invalid packet and increasing the length of other gaps. The overall rate still reaches the expected average with this technique, i.e., the accuracy is high but the precision is relatively low[3] for these delays.

A possible work-around for gaps with a length between 1 and 75 bytes is using multiple NICs to generate traffic that is sent to a switch. The switch then drops the invalid frames and multiplexes the different streams before forward-ing them to the DuT. This only works if the generated pattern can be split into multiple streams, e.g., by overlaying several Poisson processes.

---
[3] Note that $\pm 30\,ns$ is still better than hardware rate control and other software solutions (cf. Section 7.3).

However, short delays are often not meaningful in modern networks. For example, the 10GBASE-T transmission standard used by most experiments for this paper operates on frames with a payload size of 3200 bits on the physical layer as defined in IEEE 802.3 Section 4 55.1.3.1 [9]. This means that any layers above the physical layer will receive multiple packets encoded in the same frame as a burst. So two back-to-back packets cannot be distinguished from two packets with a gap of 232 bytes (185.6 ns) in the worst case and failure to represent gaps between 1 and 75 bytes should not be noticeable. Note that this limit on the physical layer only applies to relatively short inter-arrival times, bad rate control generating bursts is still inferior to our approach (cf. Figure 7 in Section 7.3).

Another limitation is that our approach is optimized for experiments in which the DuT (or the first hop in a system under test) is a software-based packet processing system and not a hardware appliance. Hardware might be affected by an invalid packet. In such a scenario, we suggest to route the test traffic through a store-and-forward switch that drops packets with invalid CRC checksums. This effectively replaces the invalid packets with real gaps on the wire. Note that the effects of the switch on the inter-arrival times need to be carefully evaluated first.

## 9. REPRODUCIBLE RESEARCH

We encourage you to install MoonGen and reproduce the results from this paper to verify our work. All experiments presented here can be reproduced with the included example scripts and NICs based on Intel 82599, X540, 82580, and XL710 chips.

The performance evaluation in Section 5 is based on the scripts found in `examples/benchmarks`, an Intel Xeon E5-2620 v3 CPU, and Intel X540 NICs. The timestamping accuracy in Section 6 was measured with the script `timestamps.lua`, the clock drift measurements with `drift.lua`. Inter-arrival times in Section 7 were measured with `inter-arrival-times.lua`. The script `l2-load-latency.lua` with the timestamping task disabled was used to generate the analyzed traffic. The suggested work-around for the hardware rate control limitations at high rates is also implemented in `l2-load-latency.lua`. Sending bursty traffic is implemented in `l2-bursts.lua`. The example measurement of the interrupt rate in Section 7.4 was conducted with `l2-load-latency.lua` and zsend 6.0.2.

`compare-rate-control-mechanisms.lua` was used for the evaluation in Section 8.2. The latency measurements with Poisson and CBR traffic in Section 8.3 are based on `l2-load-latency.lua` and `l2-poisson-load-latency.lua`. The DuT for these tests was Open vSwitch 2.0.0 on Debian Linux 3.7 with ixgbe 3.14.5 running on a server with a 3.3 GHz Intel Xeon E3-1230 v2 CPU. Only a single CPU core was used by configuring the NIC with only one queue. Each test was run for at least 30 seconds with at least 30 000 timestamped packets.

All measurements were conducted on Intel X540 NICs except for the inter-arrival times (Intel 82580), fiber loopback measurements (Intel 82599), and 40 GbE tests (Intel XL710). We used different development versions of MoonGen for the experiments described throughout this paper.

The performance evaluation with 10 GbE in Section 5 was done with commit `492c0e4`, and on 40 GbE with commit `a70ca21`. We confirmed that all described experiments still work with the example scripts from commit `a70ca21` in our git repository [5].

## 10. CONCLUSIONS AND FUTURE WORK

We have presented a general-purpose packet generator that uses hardware features of commodity NICs to implement functionality that was previously only available on expensive special-purpose hardware. MoonGen represents a hybrid between a pure software-based solution and one based on hardware. It combines the advantages of both approaches while mitigating shortcomings by using both hardware-specific features and novel software approaches.

MoonGen measures latencies with sub-microsecond accuracy and precision. The desired packet rate can be controlled precisely through both hardware-support and our rate control algorithm based on filling gaps with invalid packets.

We have shown that it is feasible to use modern implementations of scripting languages to craft packets without sacrificing speed. This makes MoonGen flexible and extensible. The flexibility goes beyond the capabilities provided by hardware load generators as each packet can be crafted in real-time by a script. Tests that respond to incoming traffic in real-time are possible as MoonGen also features packet reception and analysis.

In the future, we will add additional example scripts and support for hardware features of more NICs. MoonGen currently comes with example scripts to handle IPv4, IPv6, UDP, TCP, ICMP, IPsec, and ARP traffic.

MoonGen's flexible architecture allows for further applications like analyzing traffic in line rate on 10 GbE networks or doing Internet-wide scans from 10 GbE uplinks. MoonGen is under active development, the latest version is available in our public git repository [5].

## Acknowledgments

This research was supported by the DFG MEMPHIS project (CA 595/5-2), the KIC EIT ICT Labs on SDN, and the EUREKA-Project SASER (01BP12300A).

We would like to thank the anonymous reviewers and our colleagues Dominik Scholz, Johannes Reifferscheid, Rainer Schönberger, Patrick Werneck, Lukas Märdian, Lukas Erlacher, and Stephan M. Günther for valuable contributions to MoonGen and this paper.